\documentclass[11pt,a4paper]{article}

\usepackage[utf8]{inputenc}
\usepackage[T1]{fontenc}
\usepackage{lmodern}
\usepackage[a4paper,margin=2.45cm]{geometry}
\usepackage{amsmath,amssymb}
\usepackage{booktabs,array,longtable,graphicx,float,placeins,calc}
\usepackage{textcomp}
\usepackage{xcolor}
\usepackage{microtype}
\usepackage{authblk}
\usepackage{enumitem}
\usepackage{indentfirst}
\usepackage{caption}
\usepackage{cite}
\usepackage[colorlinks=true,linkcolor=blue,citecolor=blue,urlcolor=blue]{hyperref}

\hypersetup{
  pdftitle={Physics-Informed Symbolic Regression for Predicting the Glass Transition Temperature of Alkali Borate Glasses},
  pdfauthor={Leonardo dos Santos Vitoria; Marcio Luis Ferreira Nascimento; Susana de Souza Lalic; Daniel Roberto Cassar},
  pdfsubject={Physics-informed symbolic regression applied to alkali borate glasses},
  pdfkeywords={glass transition temperature, symbolic regression, alkali borate glasses, dissociation energy, physics-informed modeling}
}

\newcommand{\real}[1]{#1}
\newcommand{\blueeqtag}[1]{\ifmmode\tag{\textcolor{blue}{#1}}\else\textcolor{blue}{(#1)}\fi}
\newcommand{\bluetabtag}[1]{\textcolor{blue}{#1}}

\makeatletter
\renewcommand{\@biblabel}[1]{\textcolor{blue}{[#1]}}
\makeatother
\newenvironment{widefigure}{\begin{figure}[!htbp]}{\end{figure}}

\title{\textbf{Physics-Informed Symbolic Regression for Predicting the Glass Transition Temperature of Alkali Borate Glasses}}

\author[1,4]{Leonardo dos Santos Vitoria\thanks{Corresponding author: \href{mailto:leonardosvtr@gmail.com}{leonardosvtr@gmail.com}}}
\author[2]{Marcio Luis Ferreira Nascimento}
\author[1]{Susana de Souza Lalic}
\author[3]{Daniel Roberto Cassar}

\affil[1]{Center of Exact Sciences and Technology, Department of Physics, Federal University of Sergipe, Sao Cristovao, Sergipe, Brazil}
\affil[2]{Vitreous Materials Lab, Department of Chemical Engineering, Polytechnic School, Federal University of Bahia, Salvador, Bahia, Brazil}
\affil[3]{Ilum School of Science, Brazilian Center for Research in Energy and Materials (CNPEM), Campinas, Sao Paulo, Brazil}
\affil[4]{Laboratory of Advanced Materials and Strategic Minerals (LMM), Department of Physics, Federal University of Lavras (UFLA), Lavras, Minas Gerais, Brazil}

\date{}

\begin{document}

\maketitle

\begin{abstract}
\noindent The glass transition temperature (\(T_{g}\)) of alkali borate glasses is strongly composition-dependent and difficult to predict from first principles due to the structural complexity of the boron network. Here, we apply physics-informed symbolic regression (combining evolutive search with physically meaningful descriptors) to derive an interpretable closed-form expression for \(T_{g}\) in the \(x\mathrm{M}_2\mathrm{O}\cdot(100-x)\mathrm{B}_2\mathrm{O}_3\) glass family, with M = Li, Na, and K and \(x\) expressed in mol\%, and subsequently extrapolate it to M = Rb and Cs. The resulting model achieves a root-mean-square error of 14--16 K while maintaining clear physical interpretability, explicitly capturing the interplay among \(T_{g}\), structural dissociation energy, and network packing. Critically, models built on the Rigid Unit Packing Fraction (RUPF) yield substantially more realistic \(T_{g}\) predictions than those using the conventional Atomic Packing Fraction (APF), as APF overestimates structural rigidity at intermediate compositions. The fitted dissociation energies are further validated against the revised Makishima--Mackenzie model, confirming that the inferred parameters are physically consistent, not merely statistically effective, within the alkali borate family. Finally, Monte Carlo uncertainty quantification reveals that prediction uncertainty is highest in the compositional regions associated with the boron anomaly, directly linking model limitations to a known structural transition in these glasses. This result highlights the potential of physics-informed symbolic regression as a transparent and interpretable alternative to black-box models for property prediction in glass systems.
\end{abstract}


\section{Introduction}\label{introduction}

Glasses are among the oldest and most versatile classes of materials known to humankind. Their applications range from conventional uses in windows, doors, and architectural components to highly specialized technological functions, such as high-refractive-index optical glasses \cite{zhang2003,ehrt1991,zhou2018,zhu2023,li2025,alqarni2025}, chemically durable, low-solubility glasses for radioactive-waste immobilization \cite{laverov2013,rautiyal2021,ojovan2025}, and thermoluminescent glasses, which are often used in dosimetry applications \cite{ezra2024,madbouly2021,szajerski2017,elkheshen2018,thumsaard2017}. This versatility largely stems from the fact that, unlike crystals, glasses do not obey strict stoichiometric rules. This freedom allows for an enormous number of possible compositional arrangements: around \(6 \times 10^{56}\) distinct compositions can be formed by combining minimum molar concentrations of 1\% with nearly 80\% of the chemical elements \cite{zanotto2004}.

A particularly important feature of these materials is the strong correlation between their composition and their properties. Even modest compositional changes may induce significant structural rearrangements that ultimately impact macroscopic behavior. Unlike crystals, glasses are not required to follow strict stoichiometric rules, which allows their properties to be tuned continuously by changing their chemical composition. This level of flexibility is difficult to reproduce in crystalline materials.

Glasses exhibit a glass transition temperature (\(T_{g}\)), which is the temperature range at which the structural relaxation time becomes comparable to the experimental observation time. At temperatures sufficiently below \(T_{g}\), the available kinetic energy is insufficient to maintain thermodynamic equilibrium, and the material is classified as a frozen liquid, \emph{i.e.}, a glass. At temperatures sufficiently above \(T_{g}\) and below the \emph{liquidus} temperature, the kinetic energy is high enough to sustain metastable thermodynamic equilibrium, and the material is classified as a supercooled liquid \cite{zanotto2017}. Crystalline materials do not exhibit a glass transition temperature, making \(T_{g}\) one of the most relevant and defining properties of glasses.

Due to its importance, numerous studies have sought to clarify the physical nature of \(T_{g}\). Among the most influential contributions, Adam and Gibbs \cite{adam1965} related cooperative relaxation in glass-forming liquids to the configurational entropy \(S_{c}\), providing a thermodynamic basis for the strong temperature dependence of the structural relaxation time \(\tau\). Kinetically, \(T_{g}\) is commonly interpreted as the temperature at which the experimental observation time \(\tau_{obs}\) becomes comparable to the average structural relaxation time of the parent supercooled liquid \cite{zanotto2017}. Consistently, an operational experimental criterion often defines \(T_{g}\) as the temperature at which the viscosity reaches approximately \(10^{12}\,\mathrm{Pa\,s}\) \cite{mauro2009a,yue2009}. Zanotto and Mauro proposed a more recent definition of glass involving \(T_{g}\) \cite{zanotto2017}.

We are currently witnessing an era of unprecedented data availability. In glass science, databases such as SciGlass and INTERGLAD and toolkits such as GlassPy \cite{cassar2023} have encouraged data-driven approaches, most of them centered on machine learning (ML) algorithms. In recent years, several studies employing ML techniques have been published, many of which focus on neural network models for predicting glass properties based on chemical composition \cite{liu2024,cassar2018,vitoria2024,ahmmad2022}. A prominent example is GlassNet \cite{cassar2023}, proposed in 2023, which was trained on over 200,000 distinct glass compositions and can simultaneously estimate over 80 properties, including \(T_{g}\), viscosity, density, and refractive index.

Despite these advances, most predictive models induced by ML algorithms remain black-box models. Such models do not clearly expose their internal rationale when making predictions in a way that is understandable to humans. This makes it difficult to interpret the physical basis underlying their predictions \cite{hassija2024}. This limitation is unsurprising given the usual tradeoff between interpretability and predictive performance. Several tools have been proposed to compensate for this lack of transparency by extracting relevant correlations between the inputs and outputs of these black-box models (an area of study known as Explainable AI, or XAI for short). One of these tools is called SHAP \cite{lundberg2017}, which provides an efficient strategy for estimating Shapley values \cite{shapley1953}, in honor of the American mathematician Lloyd Stowell Shapley (1923--2016).

To overcome the limitations of black-box models, an alternative strategy is to combine physicochemical descriptors with symbolic regression (SR). This choice is motivated by recent studies showing that, even in purely statistical models, variables with clear physical meaning can significantly improve predictive performance \cite{shih2022,yang2025,krishnan2018,bodker2022}. SR can induce explicit mathematical expressions directly from data, in a manner analogous to the procedure used by Johannes Kepler (1571--1630) to describe planetary orbits from the observations of Tycho Brahe (1546--1601). This strategy yields algebraic models that are often inherently interpretable and capable of revealing functional relationships among variables \cite{tan2022}. In the present study, we applied SR combined with physicochemical descriptors to glasses in the \(x\mathrm{M}_2\mathrm{O}\cdot(100-x)\mathrm{B}_2\mathrm{O}_3\) family, with M = Li, Na, K, Rb, and Cs, to derive a physically interpretable and statistically accurate model for \(T_{g}\).

\section{Methodology}\label{methodology}

\subsection{Data extraction and training/validation strategy}\label{data-extraction-and-trainingvalidation-strategy}

The dataset used in this study consists of compositions of the form \(x\mathrm{M}_2\mathrm{O}\cdot(100-x)\mathrm{B}_2\mathrm{O}_3\), with \(x\) expressed in mol\%, obtained from the SciGlass database and the literature \cite{saddeek2004,ostergaard2020}. Prior to modeling, the dataset underwent a curation step aimed at removing redundant compositions.

An important precaution when developing an ML model is ensuring that no data points present in the training set also appear in the test set. Otherwise, performance metrics would be unrealistically optimistic, a phenomenon known as data leakage \cite{kaufman2012}. To prevent this, chemically equivalent compositions were grouped by atomic fractions rounded to two decimal places; repeated entries were then collapsed into a single representative record, and the median of the associated \(T_{g}\) values was assigned to that composition. The median was chosen because it is less sensitive to extreme values, helping to preserve the statistical integrity of the dataset \cite{cassar2023}.

A total of \(N_{data}\text{=}309\) compositions containing exclusively Li\textsubscript{2}O, Na\textsubscript{2}O, and K\textsubscript{2}O were selected and used to train the symbolic regression model after this process. This set was randomly divided into training (75\%) and test (25\%) subsets to evaluate the model's stability within its own chemical domain. Compositions containing Rb\textsubscript{2}O and Cs\textsubscript{2}O (\(N_{data}\text{=}142\)) were excluded from this stage and reserved for subsequent extrapolation tests. Table~\bluetabtag{1} provides a description of the studied data.

{
\begin{longtable}[]{@{}
  >{\centering\arraybackslash}p{(\linewidth - 8\tabcolsep) * \real{0.1416}}
  >{\centering\arraybackslash}p{(\linewidth - 8\tabcolsep) * \real{0.2693}}
  >{\centering\arraybackslash}p{(\linewidth - 8\tabcolsep) * \real{0.0967}}
  >{\centering\arraybackslash}p{(\linewidth - 8\tabcolsep) * \real{0.1472}}
  >{\centering\arraybackslash}p{(\linewidth - 8\tabcolsep) * \real{0.1517}}@{}}
\caption{Summary of the data as a function of the composition \(x\mathrm{M}_2\mathrm{O}\cdot(100-x)\mathrm{B}_2\mathrm{O}_3\) and the corresponding \(T_{g}\) ranges, collected from SciGlass and Ref.~\cite{berryman2001} and used in this work. For all systems, the minimum concentration \(x\) is taken as zero, corresponding to the pure boron oxide limit.}\tabularnewline
\toprule\noalign{}
\begin{minipage}[b]{\linewidth}\centering
Alkali (\(M\))
\end{minipage} & \begin{minipage}[b]{\linewidth}\centering
Number of data points
\end{minipage} & \begin{minipage}[b]{\linewidth}\centering
Max \(x\)
\end{minipage} & \begin{minipage}[b]{\linewidth}\centering
Min \(T_{g}\) {[}K{]}
\end{minipage} & \begin{minipage}[b]{\linewidth}\centering
Max \(T_{g}\) {[}K{]}
\end{minipage} \\
\midrule\noalign{}
\endhead
\bottomrule\noalign{}
\endlastfoot
Li & 125 & 83.00 & 453 & 775 \\
Na & 122 & 75.00 & 470 & 754 \\
K & 62 & 83.00 & 485 & 723 \\
Rb & 65 & 80.00 & 497 & 723 \\
Cs & 77 & 78.26 & 462 & 708 \\
\end{longtable}
}

\subsection{Training features}\label{training-features}

To induce a symbolic regression model, we assembled a set of features based on the fundamental physicochemical properties of the glasses under study. The central idea was to ensure that each input variable had a clear physical meaning, consistent with the known structural and bonding features of borate glasses. The configuration of these features was primarily based on the Makishima--Mackenzie model \cite{makishima1973}, which relates Young's modulus of oxide glasses to the dissociation energy of their structural units. We considered the refinements proposed by Shi \emph{et al.} \cite{shi2020}, who introduced a more appropriate description of atomic packing through the Rigid Unit Packing Fraction (RUPF), in contrast to the conventional Atomic Packing Fraction (APF).

Furthermore, recent studies indicate that ML models tend to perform better when their input features are derived from physically motivated quantities \cite{shih2022,yang2025,krishnan2018,bodker2022}. Motivated by this perspective, the goal of this work was to induce a model that could accurately predict experimental \(T_{g}\) values and uncover potential relationships among properties that are typically treated independently and not obviously correlated, such as dissociation energy, structural packing metrics, and the glass transition temperature.

The complete set of features was defined as a function of the molar fraction \(x\) of the modifying oxide. A detailed mathematical description is provided in the Supplementary Material, and an overview of all features and their definitions is presented in Table~\bluetabtag{2}.

{
\begin{longtable}[]{@{}
  >{\raggedright\arraybackslash}p{(\linewidth - 6\tabcolsep) * \real{0.1059}}
  >{\raggedright\arraybackslash}p{(\linewidth - 6\tabcolsep) * \real{0.3893}}
  >{\raggedright\arraybackslash}p{(\linewidth - 6\tabcolsep) * \real{0.1059}}
  >{\raggedright\arraybackslash}p{(\linewidth - 6\tabcolsep) * \real{0.3988}}@{}}
\caption{Complete set of features used in the SR model induction. Mathematical definitions are provided in the Supplementary Material.}\tabularnewline
\toprule\noalign{}
\begin{minipage}[b]{\linewidth}\raggedright
Feature
\end{minipage} & \begin{minipage}[b]{\linewidth}\raggedright
Definition
\end{minipage} & \begin{minipage}[b]{\linewidth}\raggedright
Feature
\end{minipage} & \begin{minipage}[b]{\linewidth}\raggedright
Definition
\end{minipage} \\
\midrule\noalign{}
\endhead
\bottomrule\noalign{}
\endlastfoot
\(x\) & Alkali molar concentration & \(X_{O}\) & Atomic fraction of oxygen \\
\(R\) & Ratio between modifier and network former & \(r_{O}\) & Atomic radius of oxygen \\
\(\rho_{V}\) & Volumetric density {[}g/cm\(^{3}\){]} & \(V_{O}\) & Oxygen volume {[}\AA\(^{3}\){]} \\
\(N_{at}\) & Number of atoms & \(\eta_{O}\) & Packing contribution due to oxygen \\
\(\rho_{A}\) & Atomic density {[}atoms/\AA\(^{3}\){]} & \(\eta_{i}\) & Packing contribution due to interstices \\
\(N_{3}\) & Fraction of three-fold coordinated boron & \(r_{M}\) & Ionic radius of the modifier {[}\AA{]} \\
\(N_{4}\) & Fraction of four-fold coordinated boron & \(\eta_{M}\) & Packing contribution due to modifiers \\
\(X_{M}\) & Atomic fraction of modifier ions & \(\eta_{T}\) & Total packing fraction \\
\(X_{\mathrm{BO}_{3}}\) & Atomic fraction of boron atoms in three-fold coordination & \(\rho_{Q}\) & Charge-carrier density {[}atoms/cm\(^{3}\){]} \\
\(X_{\mathrm{BO}_{4}}\) & Atomic fraction of boron atoms in four-fold coordination & \(\rho_{M}\) & Modifier density \\
\(\rho_{NBO}\) & NBO density & \(\nu\) & Poisson ratio \\
\(\rho_{N_{4}}\) & Density of four-fold coordinated boron & \(G\) & Shear modulus {[}GPa{]} \\
\(V_{NF}\) & Network-former volume {[}\AA\(^{3}\){]} & \(\sum g_{i}X_{i}\) & Atomic-fraction-weighted mean dissociation energy {[}kJ/mol{]} \\
\(V_{F}\) & Free volume for ions {[}\AA\(^{3}\){]} & \(V_{m}\) & Molar volume {[}cm\(^{3}\)/mol{]} \\
\(F^{1/3}\) & Free length for ions {[}\AA{]} & & \\
\end{longtable}
}

\subsection{Extrapolation analysis of the induced model}\label{extrapolation-analysis-of-the-induced-model}

In addition to testing within the training chemical domain, the extrapolation capability of the model was evaluated for chemically different compositions, specifically those containing Rb\textsubscript{2}O and Cs\textsubscript{2}O. However, the corresponding dissociation energies for these oxides were not available, preventing their direct use in the final induced equation.

This limitation arises because the induced model, as discussed below, is a function that depends on a subset of the studied features, including the dissociation energy \(g_{M_{2}O}\), as shown in Eq.~\blueeqtag{1}.

\[T_{g,i}^{\text{pred}}\text{=}f\left( C_{i,1},C_{i,2},\ldots,g_{M_{2}O},\ldots,C_{i,K} \right).\blueeqtag{1}\]

To overcome this limitation, the missing dissociation energies were treated as adjustable parameters following a procedure similar to that adopted by Shi \emph{et al.} \cite{shi2020}. They were determined by minimizing the mean squared error (MSE) between the experimental and predicted values:

\[g_{M_{2}O}^{fit}\text{=}\arg\min_{g_{M_{2}O}}\frac{1}{N_{fit}}\sum_{i\text{=}1}^{N_{fit}}\left( T_{g,i}^{\text{exp}}\text{-}T_{g,i}^{\text{pred}}(M_{2}O) \right)^{2}.\blueeqtag{2}\]
Here, \(N_{fit}\) denotes the number of compositions used to fit the missing dissociation energy for the extrapolation system under consideration.

No mathematical restrictions were imposed on the optimization process, since physical coherence was evaluated \emph{a posteriori}. Our hypothesis is that if the induced model captures physically consistent relationships, then the fitted energy values should satisfy the expected order of interaction strength, namely:

\(g_{{Li}_{2}O}\text{ > }g_{{Na}_{2}O}\text{ > }g_{K_{2}O}\gtrsim g_{{Rb}_{2}O}\gtrsim g_{{Cs}_{2}O}\).

To ensure a strictly independent extrapolation test, the compositions containing \({Rb}_{2}O\) and Cs\textsubscript{2}O were divided into two non-overlapping subsets. The SciGlass entries were only used for fitting \(g_{{Rb}_{2}O}\) and \(g_{{Cs}_{2}O}\), while the literature data were excluded from the fitting step and used exclusively for external testing. Thus, no composition employed to estimate the missing dissociation energies was used to evaluate their predictive performance. Model performance was evaluated using the same metrics adopted above, namely MAE and RMSE\@.

A central part of this study was assessing whether the fitted dissociation energies remained meaningful beyond the original model. For this purpose, we inserted the fitted parameters into the revised Makishima--Mackenzie (MM) model proposed by Shi \emph{et al.} \cite{shi2020}. Because Young's modulus was not used in training or parameter fitting, agreement with this second model provides stronger evidence of physical transferability than predictive accuracy alone.

For this analysis, experimental data on Young's modulus, density, and the fraction of four-fold coordinated boron units \(N_{4}\) were collected from the literature. Only compositions containing the variables required for the calculation were selected. Based on this information, Young's modulus was estimated using the model of Shi \emph{et al.} \cite{shi2020} and then compared with the corresponding experimental values using the same statistical metrics.

\subsection{Uncertainty propagation by the Monte Carlo method}\label{uncertainty-propagation-by-the-monte-carlo-method}

The glass transition temperature is often treated as a single, well-defined value for a given composition. In practice, however, \(T_{g}\) is better described as a transition interval between the supercooled liquid and glassy states \cite{zanotto2017}. This behavior is typically seen in volume--temperature and entropy--temperature cooling curves. Combined with experimental factors such as thermal history, thermal-analysis method (e.g., DSC or DTA), and cooling rate, it contributes to the natural scatter in reported values for the same composition \cite{moynihan1974,rahman2007}.

To capture this variability, the predictive uncertainty of the model was estimated with the Monte Carlo method (MC) \cite{zhang2021,papadopoulos2001}. This approach is especially suitable for nonlinear and multidimensional models, for which traditional analytical methods become impractical. For example, uncertainty propagation using the Gauss formula

\[\sigma_{T_{g}}^{2}\text{=}\sum_{i\text{=}1}^{K}\left( \frac{\partial f}{\partial C_{i}}\sigma_{C_{i}} \right)^{2}\blueeqtag{3}\]
assumes symmetric errors and independent variables, and it only applies to sufficiently small uncertainties.

In the final induced equation, the composition-dependent atomic fractions \(X_{i}\) were treated deterministically for each composition, whereas the dissociation-energy terms \(g_{i}\) were modeled as normal random variables:

\[g_{i} \sim N\left( \mu_{g_{i}},\sigma_{g_{i}}^{2} \right),\blueeqtag{4}\]
where \(\mu_{g_{i}}\) and \(\sigma_{g_{i}}\) are, respectively, the mean and standard deviation associated with each dissociation-energy term. Then, \(N_{MC}\text{=}10,000\) independent samples of the energetic parameter vector were generated:

\[g^{(j)}\text{=}\left\{ g_{1}^{(j)},g_{2}^{(j)},\ldots,g_{K}^{(j)} \right\},\quad j\text{=}1,2,\ldots,N_{MC}.\blueeqtag{5}\]
For each sampled vector, the prediction of \(T_{g}\) is given by:

\[T_{g}^{(j)}\text{=}f\left( g^{(j)},X_{i} \right).\blueeqtag{6}\]
From the simulated distribution \(\{ T_{g}^{(j)}\}_{j\text{=}1}^{N_{MC}}\), the Fisher--Pearson skewness test was applied, followed by binary classification into symmetric distributions when \(\text{|}\alpha\text{|<}0.5\) and asymmetric distributions otherwise \cite{virtanen2020}. For symmetric distributions, the final uncertainty associated with the prediction is computed as follows:

\[\sigma_{T_{g}} = \sqrt{\frac{1}{N_{\mathrm{MC}} - 1}\sum_{j=1}^{N_{\mathrm{MC}}}\left(T_{g}^{(j)} - \overline{T_{g}}\right)^{2}},\blueeqtag{7}\]
where the mean simulated value is:

\[\overline{T_{g}}\text{=}\frac{1}{N_{MC}}\sum_{j\text{=}1}^{N_{MC}}T_{g}^{(j)}.\blueeqtag{8}\]

This procedure makes it possible to estimate the propagation of uncertainty from the uncertain energetic parameters to the model output without requiring analytical derivatives. The uncertainties \(\sigma_{g_{i}}\) were estimated from the variability of the dissociation energies of the structural units reported in the literature.

\subsection{Symbolic regression algorithm}\label{symbolic-regression-algorithm}

The \(T_{g}\) model as a function of the input features \(\{ C_{1},C_{2},\ldots,C_{K}\}\) was induced using symbolic regression with the Python library PySR \cite{cranmer2023}. This tool combines evolutionary search algorithms with selection heuristics to find mathematical expressions from data.

During the induction process, the following constraints were imposed on the evolutionary search:

\begin{itemize}
\item
  A parsimony factor of 1.5 was adopted to favor compact and interpretable expressions. This parameter controls the complexity of the induced equations. In PySR, this complexity is related to the size of the symbolic expression tree, \emph{i.e.}, the number of nodes, operators, and constants used to build the equation.
\item
  The allowed operators were restricted to \(O\text{=}\{\text{+},\text{-},\text{/},\times,\exp,\log,\sqrt{\cdot}\}\).
\item
  The population consisted of 200 individuals that evolved over 1000 generations.
\end{itemize}

The search for equations was carried out without imposing dimensional-consistency constraints in PySR, allowing the algorithm to explore different functional combinations of the variables freely. The resulting expressions were then manually revised to ensure dimensional consistency and physical plausibility. In this step, dimensionless constants were reinterpreted in terms of feature sets and known physical quantities, restoring dimensional homogeneity to the final equation.

\section{Results and Discussion}\label{results-and-discussion}

We obtained the expression shown in Eq.~\blueeqtag{9} after the model selection procedure described in the methodology.

\[T_{g}\text{=}\left\lbrack 280\text{-}\left( \sum g_{i}X_{i}\text{-}222 \right)\left( X_{M_{2}O}\text{-}0.478 \right) \right\rbrack \cdot 1.72.\blueeqtag{9}\]

This raw PySR expression was examined to recover the physical meaning of its numerical constants without changing its functional form. Instead, the constants were recast in terms of physically meaningful quantities to restore dimensional consistency and physical interpretability. Thus, symbolic regression identified the functional dependence, and the subsequent (human) analysis assigned physically meaningful interpretations to its coefficients. To convert Eq.~\blueeqtag{9} into a physically interpretable and dimensionally consistent expression, the following reinterpretations were introduced:

\begin{itemize}
\item
  the constant 280 was rewritten as \(g_{\mathrm{BO}_{4}}/12\) {[}kJ mol\(^{- 1}\){]};
\item
  the constant 222 was rewritten as \(g_{\mathrm{BO}_{3}}/3\) {[}kJ mol\(^{- 1}\){]};
\item
  the factor \(X_{M_{2}O} - 0.478\) was kept dimensionless, where \(X_{M_{2}O} \equiv X_{M}\) and \(0.478\) corresponds to the packing fraction of pure borate at \(x = 0\) according to Ref.~\cite{shi2020}. We defined:
\end{itemize}

\[\Delta_{M} = X_{M_{2}O} - \eta(x = 0),\]
which measures the change in alkali content relative to the alkali-free reference state;

\begin{itemize}
\item
  the constant 1.72 was interpreted as \(\frac{23}{11}\frac{T_{g}^{\mathrm{B}_{2}\mathrm{O}_{3}}(x = 0)}{g_{\mathrm{BO}_{3}}}\), where \(T_{g}^{\mathrm{B}_{2}\mathrm{O}_{3}}(x = 0) = 540\) K \cite{mauro2009b}.
\end{itemize}

Applying these conditions, the equation for the transition temperature in kelvin becomes:

\[T_{g} = \frac{23}{11}\frac{T_{g}^{\mathrm{B}_{2}\mathrm{O}_{3}}(x = 0)}{g_{\mathrm{BO}_{3}}}\left\lbrack \frac{g_{\mathrm{BO}_{4}}}{12} - \left( \sum g_{i}X_{i} - \frac{g_{\mathrm{BO}_{3}}}{3} \right)\Delta_{M} \right\rbrack.\blueeqtag{10}\]
In Eq.~\blueeqtag{10}, \(X_{i}\) denotes the atomic fraction of boron atoms in three-fold and four-fold coordination, as well as the atomic fraction of modifiers present in the glass network (for more information, see the Supplementary Material). The constants \(g_{\mathrm{BO}_{4}}\) and \(g_{\mathrm{BO}_{3}}\) are the dissociation energies per mole associated with the borate structural groups in four-fold and three-fold coordination, respectively.

Equation~\blueeqtag{10} describes the glass transition temperature over a wide concentration range and for different alkalis using parameters predominantly associated with the dissociation energies of the structural constituents. Although PySR had access to the full feature set, including packing fraction, molar volume, and other variables, the induced model converged to a functional form dominated by dissociation-energy terms. This result suggests that dissociation energy is not only a useful predictor but also a prominent descriptor in the recovered expression. The dominance of dissociation-energy terms in the final expression is consistent with the physically plausible interpretation that local bonding strength influences \(T_{g}\).

\subsection{Evaluation of estimated dissociation energies}\label{evaluation-of-estimated-dissociation-energies}

The dissociation energies for each oxide were determined by minimizing the MSE according to Eq.~\blueeqtag{2}. In this sense, the induced symbolic model operates not only as a predictor of \(T_{g}\) but also as a physically informed proxy for estimating dissociation energies based on its algebraic structure. This feature is particularly relevant because it allows secondary physical quantities, which are not directly induced by the regression itself, to be calibrated from the functional dependence recovered by the model and subsequently examined against independent physical equations. The resulting fitted values are shown in Table~\bluetabtag{3}.

{
\begin{longtable}[]{@{}
  >{\raggedright\arraybackslash}p{(\linewidth - 14\tabcolsep) * \real{0.2569}}
  >{\centering\arraybackslash}p{(\linewidth - 14\tabcolsep) * \real{0.1293}}
  >{\centering\arraybackslash}p{(\linewidth - 14\tabcolsep) * \real{0.1034}}
  >{\centering\arraybackslash}p{(\linewidth - 14\tabcolsep) * \real{0.1161}}
  >{\centering\arraybackslash}p{(\linewidth - 14\tabcolsep) * \real{0.0823}}
  >{\centering\arraybackslash}p{(\linewidth - 14\tabcolsep) * \real{0.0798}}
  >{\centering\arraybackslash}p{(\linewidth - 14\tabcolsep) * \real{0.1034}}
  >{\centering\arraybackslash}p{(\linewidth - 14\tabcolsep) * \real{0.1288}}@{}}
\caption{Dissociation energies compiled from Refs.~\cite{makishima1973,shi2020} and estimated in this work according to Eq.~\blueeqtag{2}, including the estimates for \({Rb}_{2}O\) and \({Cs}_{2}O\). Values in parentheses are errors reported by Shi \emph{et al.} \cite{shi2020}.}\tabularnewline
\toprule\noalign{}
\begin{minipage}[b]{\linewidth}\raggedright
\end{minipage} & \multicolumn{7}{>{\centering\arraybackslash}p{(\linewidth - 14\tabcolsep) * \real{0.7431} + 12\tabcolsep}@{}}{\begin{minipage}[b]{\linewidth}\centering
\(g_{i}\) {[}kJ/mol{]}
\end{minipage}} \\
\cmidrule(l){2-8}
Source / Modifier & Li\textsubscript{2}O & Na\textsubscript{2}O & K\textsubscript{2}O & Rb\textsubscript{2}O & Cs\textsubscript{2}O & BO\textsubscript{3} & BO\textsubscript{4} \\
\midrule\noalign{}
\endhead
\bottomrule\noalign{}
\endlastfoot
Shi \emph{et al.} RUPF \cite{shi2020} & 1372(129) & 962(60) & 493(190) & --- & --- & 655(50) & 3382(121) \\
Shi \emph{et al.} APF \cite{shi2020} & 1335(140) & 836(65) & 227(207) & --- & --- & 575(54) & 3280(131) \\
Calculated from Eq.~\blueeqtag{10} & 1255 & 1003 & 488 & 452 & 431 & 676 & 3289 \\
Sun--Huggins \cite{huggins1946,sun1947} & 1205 & 1004 & 962 & 962 & 954 & --- & --- \\
\midrule\noalign{}
RUPF M--O bond length {[}\AA{]} & 1.94 & 2.37 & 2.90 & 2.95 & 3.04 & --- & --- \\
\end{longtable}
}

The fitted values for Li\textsubscript{2}O, Na\textsubscript{2}O, and K\textsubscript{2}O show only minor deviations from the average reference values reported by Shi \emph{et al.} \cite{shi2020}, while remaining within the associated uncertainty intervals. Across all oxides, particularly Rb\textsubscript{2}O and Cs\textsubscript{2}O, the fitted values follow the physically expected decreasing trend:

\[g_{{Li}_{2}O}\text{>}g_{{Na}_{2}O}\text{>}g_{K_{2}O} \gtrsim g_{{Rb}_{2}O} \gtrsim g_{{Cs}_{2}O},\]
which is consistent with the monotonic decrease and saturation behavior proposed in the additive model of Sun and Huggins \cite{huggins1946,sun1947}.

\subsection{Analysis of the induced model}\label{analysis-of-the-induced-model}

Figure~\ref{fig:packing-comparison} compares the predicted and experimental values of the glass transition temperature using the two packing-fraction formalisms: the RUPF model proposed by Shi \emph{et al.} \cite{shi2020} and the traditional APF model.

\begin{widefigure}
\centering
\caption{Comparison between values predicted by the model and experimental values. This comparison considers both the test set for Li, Na, and K and the data reserved exclusively for the extrapolation stage. The left panel corresponds to the model using RUPF, whereas the right panel corresponds to the APF model.}
\label{fig:packing-comparison}
\includegraphics[width=\textwidth]{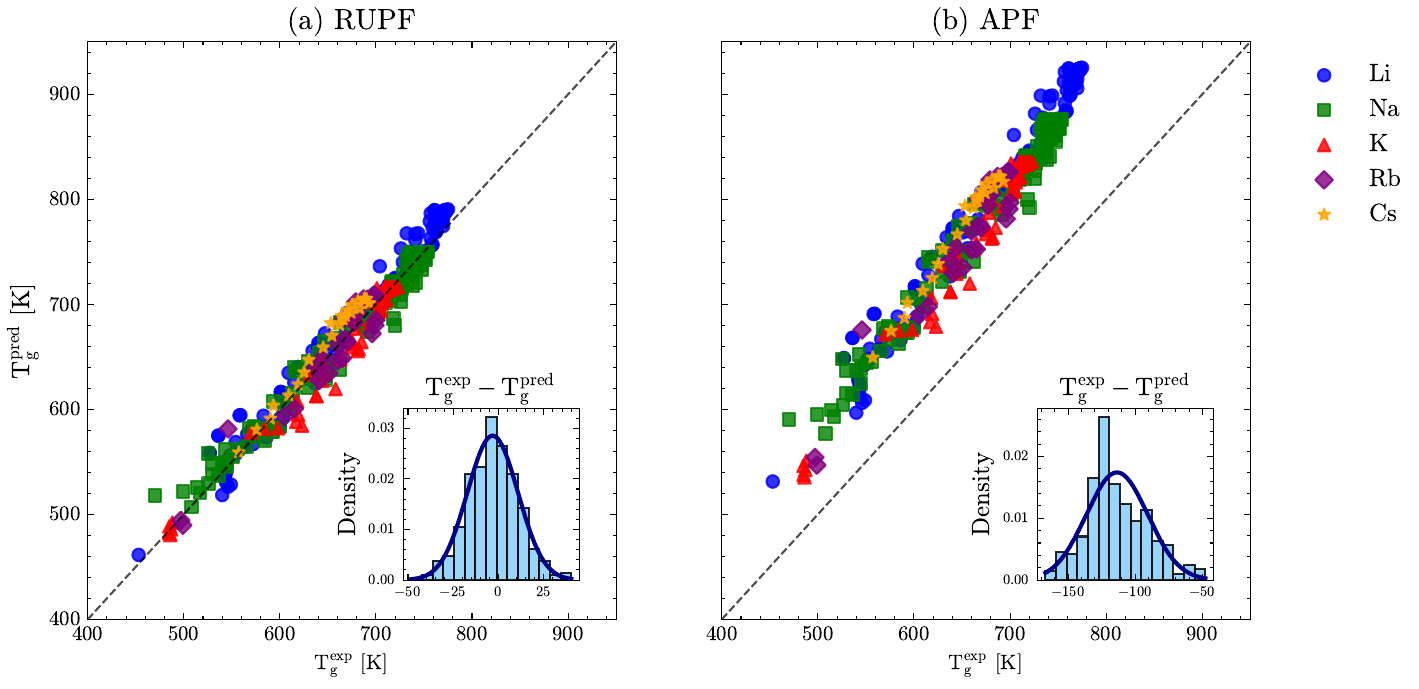}
\end{widefigure}

The dashed diagonal lines in Figure~\ref{fig:packing-comparison} represent the ideal fit region, that is, the condition \(T_{g}^{\text{pred}}\text{=}T_{g}^{\text{exp}}\). The side and bottom insets correspond to the residual distributions. Statistically, the residuals of a good inference model should follow a normal distribution with a mean close to zero, indicating an absence of statistical bias.

Table~\bluetabtag{4} summarizes the main statistical metrics for the test and extrapolation sets under the two packing-fraction models.

\begin{table}[H]
\centering
\caption{Statistical metrics evaluated on the test set (Li, Na, and K) and the extrapolation set (Rb and Cs), comparing the two packing-fraction models.}
\begin{tabular}{lcccc}
\toprule
& \multicolumn{2}{c}{RUPF} & \multicolumn{2}{c}{APF} \\
\cmidrule(lr){2-3}\cmidrule(lr){4-5}
System/Metrics & MAE [K] & RMSE [K] & MAE [K] & RMSE [K] \\
\midrule
Li      & 11.37 & 13.78 &  96.65 &  99.77 \\
Na      & 14.39 & 17.49 &  78.53 &  80.20 \\
K       & 11.76 & 15.50 &  70.21 &  72.56 \\
Rb      &  9.70 & 12.48 &  99.44 & 100.87 \\
Cs      & 15.03 & 17.22 & 106.71 & 109.28 \\
Overall & 12.61 & 15.61 &  86.70 &  89.86 \\
\bottomrule
\end{tabular}
\end{table}

According to Table~\bluetabtag{4}, the induced model exhibits good overall statistical performance when the RUPF packing fraction is employed. The overall RMSE of 15.61 K and the MAE of 12.61 K support this conclusion. Moreover, the local RMSE values range from 12.48 to 17.49 K. This shows that, although the model was trained only on glasses containing Li\textsubscript{2}O, Na\textsubscript{2}O, and K\textsubscript{2}O, it can still extrapolate satisfactorily to compositions outside the training chemical domain. The accuracy is reasonably preserved in the considered extrapolation setting. This extrapolative behavior is noteworthy in the present setting and indicates that the physically informed input variables capture key trends in composition, structure, and properties relevant to this dataset.

Furthermore, the \(\mathrm{Rb}_2\mathrm{O}\)-containing system exhibited the lowest MAE among the extrapolation cases (9.70 K) on the external literature subset, which was not used to fit \(g_{Rb_{2}O}\). Because the induced model was trained using alkalis with smaller ionic radii and, therefore, lower polarizability, one might expect it to overestimate the transition temperature for larger ions, such as \(\mathrm{Rb}^{+}\) and \(\mathrm{Cs}^{+}\), which interact less strongly with the glass former \(\mathrm{B}_2\mathrm{O}_3\) than smaller ions such as \(\mathrm{Li}^{+}\) \cite{wu2014,kojima2020}. However, this trend is not observed. We hypothesize that this behavior is related to the inclusion of dissociation energies as a feature. These energies are associated with the complete dissociation of structural units in the network and, in a broad sense, correlate with ionic size and bond strength \cite{langhoff1986,dolizy1986}. Thus, the model was able to capture this intrinsic physical dependence without explicitly including ionic radius as an input variable.

In contrast, as illustrated in Figure~\ref{fig:packing-comparison}, the conventional APF model leads to systematic errors when predicting \(T_{g}\). For pure B\textsubscript{2}O\textsubscript{3} (\(x\text{=}0\)), Shi \emph{et al.} \cite{shi2020} reported an RUPF value of 0.47, whereas the APF assumes a larger value of 0.49. This overestimation by the APF produces a less realistic representation of the structural compactness of BO\textsubscript{3} units, which in turn leads to an overestimation of the network rigidity and, consequently, \(T_{g}\). The superiority of the RUPF model is not merely statistical. By providing a more realistic packing fraction, RUPF avoids the systematic rigidity overestimation introduced by the APF\@. Consequently, it yields a more realistic \(T_{g}\) response across compositions. This improvement is consistent with the physical relevance of the packing formalism embedded in the model.

\subsection{Interpretation of the induced model}\label{interpretation-of-the-induced-model}

Beyond standard statistical analysis, the induced model also enables qualitative interpretations of the glass network and the thermodynamics of the glass transition. From a structural standpoint, the model highlights the importance of network connectivity and rigidity: the addition of more \(\mathrm{BO}_4\) units implies a greater number of B--O--B linkages that strengthen the structure. According to topological constraint theory, hyperstatic networks with fewer floppy modes tend to exhibit higher glass transition temperatures \cite{mauro2009b,mauro2011}.

In the present model, the contribution of four-fold coordinated units is represented by \(g_{\mathrm{BO}_{4}}\text{/}12\), which defines a reference value associated with the intrinsic rigidity of this bond topology. Conversely, the term \(\left\lbrack \sum g_{i}X_{i}\text{-}\frac{g_{\mathrm{BO}_{3}}}{3} \right\rbrack\Delta_{M}\) modulates the increase in rigidity by adjusting the influence of \(\mathrm{BO}_{3}\) units. This control is linked to the initial structure of pure borate, whose connectivity and structural packing serve as the reference for the limiting case of no modifiers (\(x\text{=}0\)). In other words, the contribution to Eq.~\blueeqtag{10} depicted above illustrates how the network\textquotesingle s behavior changes as modifier ions are incorporated.

Thus, as the concentration of M\textsubscript{2}O increases, \(T_{g}\) initially rises, reflecting the formation of additional BO\textsubscript{4} units and the corresponding increase in packing. However, above roughly 30 mol\% of M\textsubscript{2}O, there are insufficient boron atoms available to generate new BO\textsubscript{4} units. The structure then begins to incorporate non-bridging oxygens (NBOs); the packing and \(T_{g}\) decrease accordingly \cite{kroghmoe1969,doweidar1990}. Thus, the well-known boron anomaly is reproduced by the induced model.

More generally, the model describes the dependence of \(T_{g}\) on the initial threefold coordination state. At \(x\text{=}0\), the network is fully polymerized by \(\mathrm{BO}_{3}\) units and undergoes a glass transition at approximately \(T_{g}(x\text{=}0) \approx 540\) K \cite{huggins1946}. Changes in composition alter boron coordination and, consequently, the network-average energy term \(\sum g_{i}X_{i}\). Because this energy term is directly proportional to the atomic fractions, it follows the boron-anomaly trend: the larger the mean dissociation energy, the stronger the interatomic bonds and the greater the structural connectivity. Therefore, the transition temperature is higher.

\subsection{Uncertainty analysis via the Monte Carlo method}\label{uncertainty-analysis-via-the-monte-carlo-method}

As discussed above, the glass transition temperature is influenced by factors that are generally difficult to control or eliminate. Even a change in the measurement technique, such as from DSC to DTA, can lead to discrepancies in reported values. This inherent variability results in a natural spread in \(T_{g}\) data, supporting the analysis of transition regions rather than relying on single-point estimates. Using the Monte Carlo method described in Section 2, we estimated the uncertainty propagated through Eq.~\blueeqtag{10}.

The first noteworthy feature of the calculated uncertainties is that they do not remain constant as the concentration of modifier ions increases. This behavior arises because Eq.~\blueeqtag{10} depends on the energetic term \(\sum g_{i}X_{i}\), which varies nonlinearly with composition through the atomic fractions \(X_{i}\). Compositions around 30\% alkali addition exhibit the largest uncertainties, as the energetic balance becomes particularly sensitive to changes in boron coordination within this range. In this interval, the contribution associated with BO\textsubscript{4} units gains importance, leading to a stronger impact of its uncertainty on the propagated bands. However, this increase should not be attributed exclusively to BO\textsubscript{4}; rather, it reflects the cooperative balance among all energetic contributions to \(\sum g_{i}X_{i}\), whose relative weights change with composition. Consequently, the uncertainty is maximized in the compositional range where the network-average energetic term is most sensitive to coordination changes.

\begin{widefigure}
\centering
\caption{Behavior of the glass transition temperature as a function of composition for (a) K, (b) Li, and (c) Na glasses. The bands around the function represent the uncertainties calculated by the Monte Carlo method using Equation~\blueeqtag{7}, i.e., the type-A standard uncertainty associated with the predicted value. The points correspond to the experimental values.}
\label{fig:uncertainty-bands}
\includegraphics[width=\linewidth]{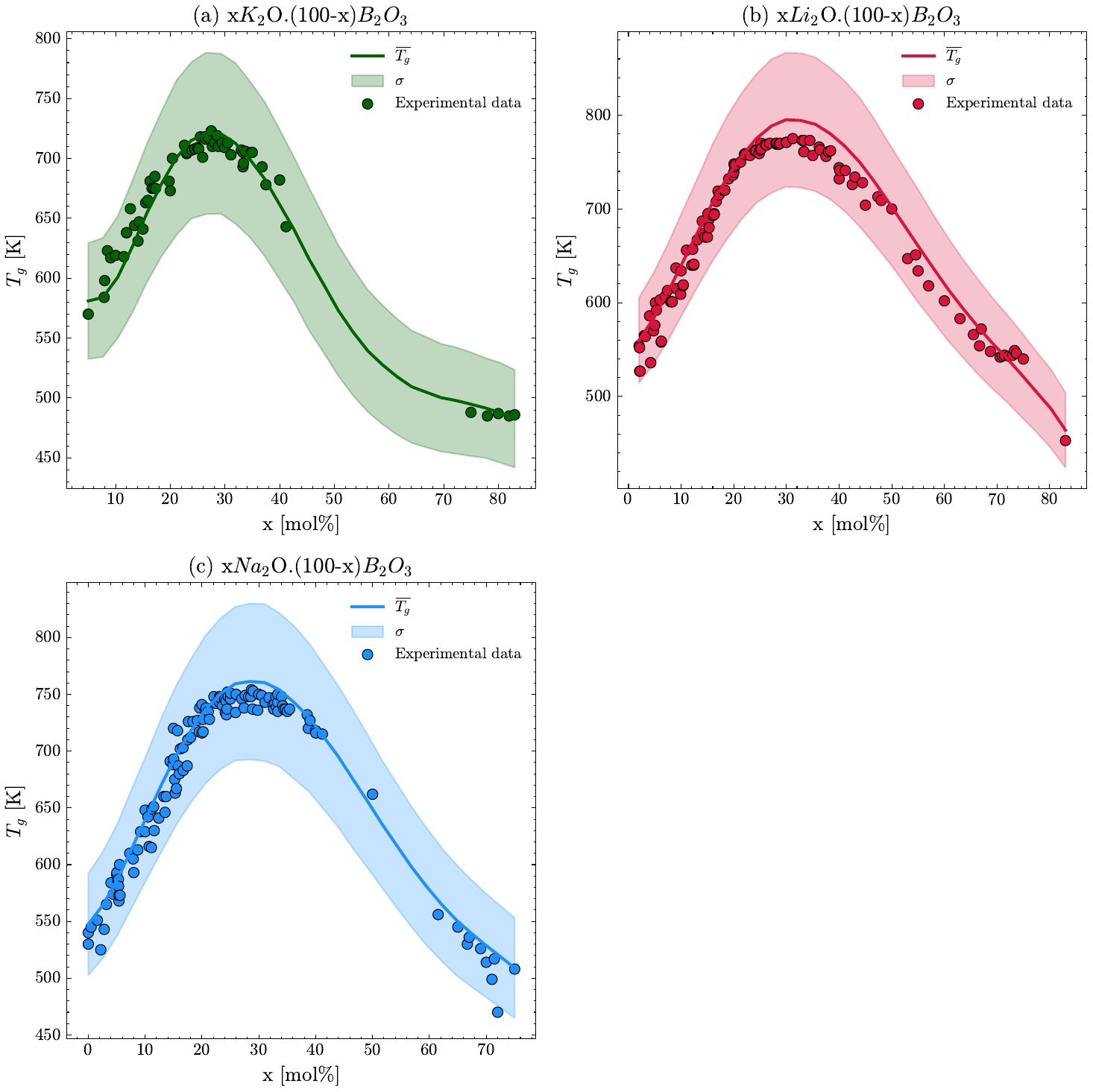}
\end{widefigure}
\FloatBarrier

\subsection{Cross-property validation with Young's modulus}\label{cross-property-validation-with-youngs-modulus}

One of the central goals of this study is to validate the fitted dissociation energies across properties within the revised MM model proposed by Shi and co-workers \cite{shi2020}. A successful consistency check would indicate that Eq.~\blueeqtag{10} captures physically meaningful relationships in the studied glass systems and that the adopted methodology may provide insight into properties that are not directly correlated \emph{a priori}.

This validation was carried out for B\textsubscript{2}O\textsubscript{3}--Rb\textsubscript{2}O and B\textsubscript{2}O\textsubscript{3}--Cs\textsubscript{2}O glasses using dissociation energies predicted by the error-minimization procedure. A final set of 47 compositions satisfied the criteria established in the literature survey described in Section 2.3. Twenty of these correspond to the B\textsubscript{2}O\textsubscript{3}--Rb\textsubscript{2}O system, and the remaining compositions correspond to the B\textsubscript{2}O\textsubscript{3}--Cs\textsubscript{2}O system. For the cesium-containing system, only five compositions have experimental Young\textquotesingle s modulus values reported in Ref.~\cite{ostergaard2020}. The remaining compositions rely on estimates from the Makishima--Mackenzie model reported in Ref.~\cite{saddeek2004}. For the rubidium-containing system, no experimental Young\textquotesingle s modulus values were found in the literature; only theoretical estimates reported in Ref.~\cite{saddeek2004} were available for comparison.

\begin{figure}[!htbp]
\centering
\caption{Predicted \(E^{pred}\) values, calculated using the model of Shi \emph{et al.} \cite{shi2020} and the dissociation energies obtained in this work for rubidium and cesium oxides, versus theoretical or experimental \(E^{\exp}\) values. The closer the points are to the central line, the better the agreement. Star markers denote experimental values reported exclusively in the literature for the cesium-containing system.}
\label{fig:young-parity}
\includegraphics[width=0.86\linewidth]{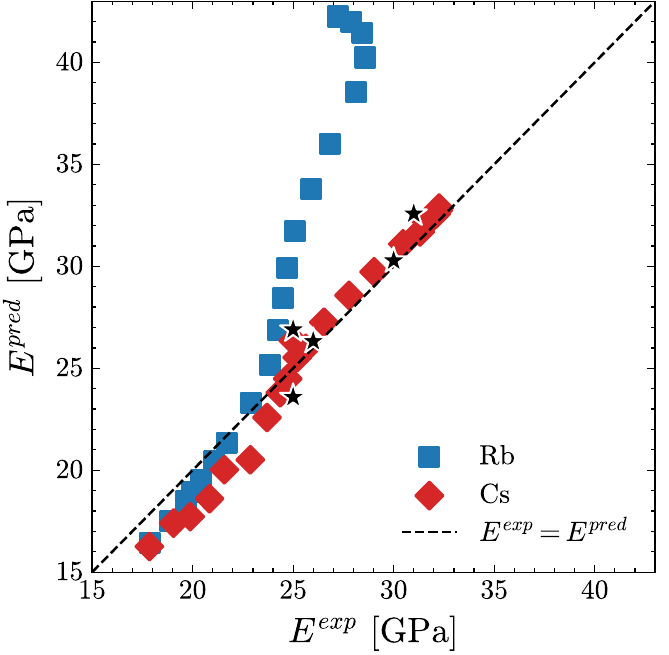}
\end{figure}

Table~\bluetabtag{5} presents a representative subset of the data plotted in Figure~\ref{fig:young-parity}. Here, \(E^{\exp}\) denotes the reference Young's modulus (experimental where available and otherwise model-derived), whereas \(E^{pred}\) denotes the Young's modulus predicted from the dissociation energies listed in Table~\bluetabtag{3}.

\begingroup
\small
\begin{longtable}[]{@{}
  >{\raggedleft\arraybackslash}p{(\linewidth - 16\tabcolsep) * \real{0.0820}}
  >{\raggedleft\arraybackslash}p{(\linewidth - 16\tabcolsep) * \real{0.0832}}
  >{\raggedleft\arraybackslash}p{(\linewidth - 16\tabcolsep) * \real{0.0808}}
  >{\raggedleft\arraybackslash}p{(\linewidth - 16\tabcolsep) * \real{0.0808}}
  >{\raggedleft\arraybackslash}p{(\linewidth - 16\tabcolsep) * \real{0.0936}}
  >{\raggedleft\arraybackslash}p{(\linewidth - 16\tabcolsep) * \real{0.0808}}
  >{\raggedleft\arraybackslash}p{(\linewidth - 16\tabcolsep) * \real{0.1504}}
  >{\raggedleft\arraybackslash}p{(\linewidth - 16\tabcolsep) * \real{0.1515}}
  >{\raggedleft\arraybackslash}p{(\linewidth - 16\tabcolsep) * \real{0.1970}}@{}}
\caption{Partial results obtained by applying the dissociation energies estimated in this work to the Young\textquotesingle s modulus model proposed in Ref.~\cite{shi2020}. GlassNet values are included as an external machine-learning reference.}\tabularnewline
\toprule\noalign{}
\begin{minipage}[b]{\linewidth}\raggedleft
B\(_{2}\)O\(_{3}\)
\end{minipage} & \begin{minipage}[b]{\linewidth}\raggedleft
Rb\(_{2}\)O
\end{minipage} & \begin{minipage}[b]{\linewidth}\raggedleft
Cs\(_{2}\)O
\end{minipage} & \begin{minipage}[b]{\linewidth}\raggedleft
\[N_{4}\]
\end{minipage} & \begin{minipage}[b]{\linewidth}\raggedleft
\[V_{m}\]
\end{minipage} & \begin{minipage}[b]{\linewidth}\raggedleft
\[\eta_{T}\]
\end{minipage} & \begin{minipage}[b]{\linewidth}\raggedleft
\(E^{\exp}\) {[}GPa{]}
\end{minipage} & \begin{minipage}[b]{\linewidth}\raggedleft
\(E^{pred}\) {[}GPa{]}
\end{minipage} & \begin{minipage}[b]{\linewidth}\raggedleft
GlassNet {[}GPa{]}
\end{minipage} \\
\midrule\noalign{}
\endfirsthead
\toprule\noalign{}
\begin{minipage}[b]{\linewidth}\raggedleft
B\(_{2}\)O\(_{3}\)
\end{minipage} & \begin{minipage}[b]{\linewidth}\raggedleft
Rb\(_{2}\)O
\end{minipage} & \begin{minipage}[b]{\linewidth}\raggedleft
Cs\(_{2}\)O
\end{minipage} & \begin{minipage}[b]{\linewidth}\raggedleft
\[N_{4}\]
\end{minipage} & \begin{minipage}[b]{\linewidth}\raggedleft
\[V_{m}\]
\end{minipage} & \begin{minipage}[b]{\linewidth}\raggedleft
\[\eta_{T}\]
\end{minipage} & \begin{minipage}[b]{\linewidth}\raggedleft
\(E^{\exp}\) {[}GPa{]}
\end{minipage} & \begin{minipage}[b]{\linewidth}\raggedleft
\(E^{pred}\) {[}GPa{]}
\end{minipage} & \begin{minipage}[b]{\linewidth}\raggedleft
GlassNet {[}GPa{]}
\end{minipage} \\
\midrule\noalign{}
\endhead
\bottomrule\noalign{}
\endlastfoot
99.04 & 0.96 & 0.00 & 0.010 & 37.314 & 0.479 & 18.88 & 17.49 & 19.20 \\
98.05 & 1.95 & 0.00 & 0.020 & 37.184 & 0.484 & 19.66 & 18.45 & 19.96 \\
97.51 & 2.49 & 0.00 & 0.025 & 37.118 & 0.487 & 19.97 & 18.92 & 20.10 \\
96.91 & 3.09 & 0.00 & 0.032 & 37.049 & 0.490 & 20.43 & 19.52 & 20.26 \\
95.95 & 4.05 & 0.00 & 0.042 & 36.948 & 0.495 & 21.08 & 20.42 & 20.50 \\
90.00 & 0.00 & 10.00 & 0.089 & 37.913 & 0.515 & 25.00 & 23.17 & 27.69 \\
85.00 & 0.00 & 15.00 & 0.108 & 38.499 & 0.531 & 26.00 & 24.15 & 27.94 \\
80.00 & 0.00 & 20.00 & 0.147 & 39.285 & 0.543 & 25.00 & 26.05 & 28.00 \\
75.00 & 0.00 & 25.00 & 0.194 & 40.233 & 0.553 & 30.00 & 28.33 & 28.06 \\
70.00 & 0.00 & 30.00 & 0.267 & 41.317 & 0.560 & 31.00 & 31.40 & 27.92 \\
\end{longtable}
\endgroup

Figure~\ref{fig:young-parity} shows the assessment of the model using Young\textquotesingle s modulus recalculated from the dissociation energies fitted in this study. For the B\textsubscript{2}O\textsubscript{3}--Cs\textsubscript{2}O system, when the five experimental values from Ref.~\cite{ostergaard2020} are combined with additional compositions estimated in Ref.~\cite{saddeek2004}, the model reproduces the trends consistently, with an MAE of 1.07 GPa and an RMSE of 1.26 GPa. These values fall within the expected prediction fluctuation of approximately 2.9 GPa reported by Shi \emph{et al.} \cite{shi2020}, supporting the physical consistency of the dissociation energies estimated here with the mechanical behavior of the system.

For the B\textsubscript{2}O\textsubscript{3}--Rb\textsubscript{2}O system, the comparison yields an MAE of 4.81 GPa and an RMSE of 6.55 GPa, which are significantly larger than the values obtained for the cesium system. These errors may partly reflect uncertainties in the theoretical reference values, since the classical Makishima--Mackenzie model tends to show systematic deviations in Young\textquotesingle s modulus due to its simplified treatment of atomic packing \cite{shi2020}. In this case, the comparison should be interpreted primarily as a test of physical consistency and order of magnitude rather than as a strict experimental validation. Nevertheless, the fact that the predicted moduli preserve the correct order of magnitude indicates that the fitted energies remain physically reasonable.

A more qualitative interpretation can also be drawn. Since the average dissociation energy term evolves similarly for the systems containing Rb\textsubscript{2}O and Cs\textsubscript{2}O over much of the compositional range, the observed differences in Young\textquotesingle s modulus are unlikely to arise from dissociation energy alone. Instead, they appear to reflect the combined effects of structural packing and molar volume. This indicates that energetic terms and packing-fraction terms play distinct but complementary roles in these mechanically related comparisons.

Taken together, these results show that the present methodology was assessed at three complementary levels: first, by extrapolation to chemically distinct alkali borate systems not included in the training set; second, through the physically reasonable ordering and magnitude of the fitted dissociation energies; and third, by evaluating these fitted parameters within the revised MM model of Shi \emph{et al.} \cite{shi2020}. In this context, the present study can be viewed both as a predictive strategy for \(T_{g}\) and as a physically guided assessment, in which symbolic induction, parameter calibration, and external property transfer jointly support the inferred structure--property relationship. Although the symbolic model was developed exclusively to describe \(T_{g}\), it captures structure--property relationships beyond \(T_{g}\) prediction and remains physically consistent under an additional cross-property validation step within the studied family.

These results suggest a potentially useful direction for future studies aimed at developing predictive models for properties with abundant data, such as \(T_{g}\) and density, expressed in terms of physical features for which data are scarce, such as dissociation energy \(g\), formation enthalpy \(H\), and fragility index \(m\). At the same time, no broader transferability should be assumed a priori, since chemically distinct glass families may be governed by different structural and energetic features. If this methodology proves informative beyond the present system through further validation, it may provide an uncertainty-aware approach for estimating physical properties that are more difficult to measure directly in other glass families. This would support more interpretable modeling of glass properties.
\FloatBarrier

\section{Conclusions}\label{conclusions}

We developed a symbolic regression model based on physicochemical descriptors to construct an explicit expression for the glass transition temperature \(T_{g}\) of glasses in the \(x\mathrm{M}_2\mathrm{O}\cdot(100-x)\mathrm{B}_2\mathrm{O}_3\) family, where M = Li, Na, K, Rb, or Cs and \(x\) is expressed in mol\%. The final model shows good statistical performance while preserving physical consistency. In addition, it extrapolates to compositions not included in the training set with reasonably maintained accuracy under the considered extrapolation conditions.

A particularly important result is that the fitted dissociation energies satisfactorily reproduce Young\textquotesingle s modulus when used with the model of Shi \emph{et al.} \cite{shi2020}. This finding reinforces the physical consistency of the method and provides an additional validation step. The uncertainty analysis based on a Monte Carlo approach allows the estimation of error propagation within the adopted uncertainty model and helps identify critical regions of variation in \(T_{g}\).

These results indicate that symbolic regression can recover physically meaningful structure-property relationships in binary alkali borate glasses while maintaining predictive accuracy and extrapolation capability within the chemical domain examined here. More broadly, the findings support the use of physically informed symbolic models as a promising route toward interpretable modeling of alkali borate glasses. Nevertheless, transfer to chemically distinct glass families should not be assumed a priori, as different systems may be governed by different balances among bonding, packing, and network-topological effects.

\section*{Acknowledgments}

The authors thank the São Paulo Research Foundation (FAPESP) for the Visiting Professor Grant no. 2025/09121-2. This study was financed in part by the Brazilian National Council for Scientific and Technological Development (CNPq), grant \#304584/2023-1. This study was partially financed by the Coordenação de Aperfeiçoamento de Pessoal de Nível Superior--Brasil (CAPES), Finance Code 001. DRC acknowledges funding from the National Institute of Science and Technology on Materials Informatics (INCT--CNPq), Campinas, Brazil (grant no. 371610/2023-0), as well as from the São Paulo Research Foundation (FAPESP) under grant nos. 2023/09820-2 (thematic project) and 2024/00989-7 (CEPID-CEMol).

\bibliographystyle{unsrt}
\bibliography{main}

\clearpage
\begin{center}
{\Large\bfseries Supplementary Material\par}
\end{center}
\vspace{0.75em}
\appendix
\section{Supplementary derivations}\label{appendix-supplementary-derivations}

This supplementary section summarizes the mathematical definitions used to build the physicochemical descriptors employed throughout the manuscript. Unless otherwise stated, the glass composition is written as \(x\mathrm{M}_2\mathrm{O}\cdot(100-x)\mathrm{B}_2\mathrm{O}_3\), where \(x\) is expressed in mol\%.

\subsection{Polynomial descriptions of density and boron coordination}\label{a.1-polynomial-descriptions-of-density-and-boron-coordination}

Two experimentally informed quantities are required as inputs for the descriptor construction: the volumetric density \(\rho_{V}(x)\) and the fraction of four-fold coordinated boron \(N_{4}(x)\). For each alkali system, both quantities were represented by composition-dependent polynomial fits:

\[\rho_{V}(x) = a_{3}x^{3} + a_{2}x^{2} + a_{1}x + a_{0},\blueeqtag{1}\]

\[N_{4}(x) = b_{5}x^{5} + b_{4}x^{4} + b_{3}x^{3} + b_{2}x^{2} + b_{1}x + b_{0},\blueeqtag{2}\]
where the coefficients \(\{ a_{i}\}\) and \(\{ b_{i}\}\) depend on the modifying oxide. The fraction of three-fold coordinated boron is then

\[N_{3}(x) = 1 - N_{4}(x).\blueeqtag{3}\]

\subsection{Auxiliary compositional quantities}\label{a.2-auxiliary-compositional-quantities}

The ratio between modifier and former oxides is written as

\[R = \frac{x}{100 - x}.\blueeqtag{4}\]

The total number of atoms in the corresponding formula unit is therefore

\[N_{at} = 3R + 5.\blueeqtag{5}\]

For a modifier with atomic mass \(m_{M}\), the mean molar mass of the glass composition is

\[M_{T}(x) = \left( \frac{x}{100} \right)\left( 2m_{M} + m_{O} \right) + \left( \frac{100 - x}{100} \right)\left( 2m_{B} + 3m_{O} \right),\blueeqtag{6}\]
where \(m_{B}\) and \(m_{O}\) are the atomic masses of boron and oxygen, respectively.

The atomic density is defined as

\[\rho_{A} = \frac{0.6022\,\rho_{V}\, N_{at}}{R(2m_{M} + m_{O}) + (2m_{B} + 3m_{O})}.\blueeqtag{7}\]

\subsection{\texorpdfstring{Atomic fractions used in the \(T_{g}\) model}{Atomic fractions used in the Tg model}}\label{a.3-atomic-fractions-used-in-the-t_g-model}

The symbolic-regression model for \(T_{g}\) uses atomic-fraction weighting. In this formulation, the atomic fraction of modifier cations is

\[X_{M} = \frac{2R}{N_{at}},\blueeqtag{8}\]
whereas the atomic fractions of boron atoms in three-fold and four-fold coordination are

\[X_{\mathrm{BO}_{3}} = \frac{2N_{3}}{N_{at}} = \frac{2(1 - N_{4})}{N_{at}},\blueeqtag{9}\]

\[X_{\mathrm{BO}_{4}} = \frac{2N_{4}}{N_{at}}.\blueeqtag{10}\]

The atomic fraction of oxygen is correspondingly

\[X_{O} = \frac{R + 3}{N_{at}}.\blueeqtag{11}\]

The mean dissociation-energy term entering the induced \(T_{g}\) equation is then

\[\sum_{i}^{}g_{i}X_{i} = g_{M_{2}O}X_{M} + g_{\mathrm{BO}_{3}}X_{\mathrm{BO}_{3}} + g_{\mathrm{BO}_{4}}X_{\mathrm{BO}_{4}}.\blueeqtag{12}\]

In the present \(T_{g}\) model, this energetic term is written in terms of atomic fractions.

\subsection{Packing-fraction descriptors}\label{a.4-packing-fraction-descriptors}

The average effective oxygen volume is written as

\[V_{O} = \frac{\frac{4\pi}{3}\left\lbrack 3X_{\mathrm{BO}_{3}}r_{\mathrm{BO}_{3}}^{3} + 4X_{\mathrm{BO}_{4}}r_{\mathrm{BO}_{4}}^{3} \right\rbrack}{3X_{\mathrm{BO}_{3}} + 4X_{\mathrm{BO}_{4}}},\blueeqtag{13}\]
where \(r_{\mathrm{BO}_{3}}\) and \(r_{\mathrm{BO}_{4}}\) are the effective oxygen radii in \(\mathrm{BO}_{3}\) and \(\mathrm{BO}_{4}\) units, respectively.

The oxygen contribution to the packing fraction is

\[\eta_{O} = V_{O}\rho_{A}X_{O}.\blueeqtag{14}\]

The interstitial contribution follows the form

\[\eta_{i} = \frac{f_{3}}{1 - f_{3}}\rho_{A}\frac{4\pi}{3}X_{\mathrm{BO}_{3}}r_{\mathrm{BO}_{3}}^{3} + \frac{f_{4}}{1 - f_{4}}\rho_{A}\frac{4\pi}{3}X_{\mathrm{BO}_{4}}r_{\mathrm{BO}_{4}}^{3},\blueeqtag{15}\]
with \(f_{3} = 0.395\) and \(f_{4} = 0.260\) for the trigonal and tetrahedral borate units, respectively.

The effective oxygen radius is obtained from

\[r_{O} = \left( \frac{3V_{O}}{4\pi} \right)^{1/3},\blueeqtag{16}\]
and the effective modifier radius is

\[r_{M} = r_{M\text{-}O} - r_{O},\blueeqtag{17}\]
where \(r_{M\text{-}O}\) is the characteristic M--O bond length.

The modifier contribution is therefore

\[\eta_{M} = \frac{4\pi}{3}r_{M}^{3}\rho_{A}X_{M},\blueeqtag{18}\]
and the total rigid-unit packing fraction becomes

\[\eta_{T} = \eta_{O} + \eta_{i} + \eta_{M}.\blueeqtag{19}\]

\subsection{Molar volume and molar-fraction formulation for Young's modulus}\label{a.5-molar-volume-and-molar-fraction-formulation-for-youngs-modulus}

The molar volume used in the Young's modulus treatment is

\[V_{m} = \frac{M_{T}(x)}{\rho_{V}(x)}.\blueeqtag{20}\]

Although the \(T_{g}\) model uses the energetic term in Eq.~\blueeqtag{12}, the Young's modulus calculation is written in terms of molar fractions of the structural units. In the Makishima--Mackenzie formulation, and likewise in the treatment discussed by Shi \emph{et al.} \cite{shi2020}, this energetic contribution is expressed in molar rather than atomic terms. To this end, the molar amounts of the structural species are defined as:

\[n_{M_{2}O} = x,\blueeqtag{21}\]

\[n_{\mathrm{BO}_{3}} = 2(100 - x)\left( 1 - N_{4} \right),\blueeqtag{22}\]

\[n_{\mathrm{BO}_{4}} = 2(100 - x)N_{4}.\blueeqtag{23}\]

The total molar amount is then

\[n_{tot} = n_{M_{2}O} + n_{\mathrm{BO}_{3}} + n_{\mathrm{BO}_{4}},\blueeqtag{24}\]
leading to the molar fractions

\[\chi_{M_{2}O} = \frac{n_{M_{2}O}}{n_{tot}},\quad\quad\chi_{\mathrm{BO}_{3}} = \frac{n_{\mathrm{BO}_{3}}}{n_{tot}},\quad\quad\chi_{\mathrm{BO}_{4}} = \frac{n_{\mathrm{BO}_{4}}}{n_{tot}}.\blueeqtag{25}\]

The corresponding mean dissociation-energy term is thus

\[\langle g\rangle_{mol} = g_{M_{2}O}\chi_{M_{2}O} + g_{\mathrm{BO}_{3}}\chi_{\mathrm{BO}_{3}} + g_{\mathrm{BO}_{4}}\chi_{\mathrm{BO}_{4}}.\blueeqtag{26}\]

Using this energetic term, the Young's modulus is written as

\[E = \frac{2\eta_{T}\langle g\rangle_{mol}}{V_{m}},\blueeqtag{27}\]
and the shear modulus follows from

\[G = \frac{E}{2(1 + \nu)},\blueeqtag{28}\]
where \(\nu\) is the Poisson ratio.

The distinction between Eqs.~\blueeqtag{12} and \blueeqtag{26} is essential in the present work: the induced \(T_{g}\) model uses atomic fractions, whereas the Young's modulus treatment follows the molar-fraction formalism adopted in the Makishima--Mackenzie model and in the later treatment of Shi \emph{et al.} \cite{shi2020}.

\end{document}